\documentclass{article}[11pt]
\usepackage{array}
\usepackage{amsmath}
\usepackage{calc}
\usepackage{epsfig}

\newcommand{\PBS}[1]{\let\temp=\\#1\let\\=\temp}
\newcommand{\Tr}{\mathop{\mathrm{Tr}}\nolimits}

\newtheorem{theorem}{Theorem}

\begin{document}

\title{Quantum Phase Space in Relativistic Theory:\\ the Case of
Charge-Invariant Observables\footnote{Oral talk given at 5th
International Conference "Symmetry in Nonlinear Mathematical
Physics", Kiev, Ukraine (June 23-29, 2003)}}

\author{\vspace{3ex} A.A. Semenov$^{1}$\footnote{E-mail:
sem@iop.kiev.ua}, B.I. Lev$^{1,2}$\footnote{E-mail:
lev@iop.kiev.ua}, C.V. Usenko$^{2}$\footnote{E-mail:
usenko@ups.kiev.ua}\\
 \parbox[c]{0.9\textwidth}{$^{1}$\textsl{\small
Institute of Physics, National Academy of Sciences of Ukraine, 46
\mbox{Nauky pr,} Kiev 03028, Ukraine}} \\
\parbox[c]{0.9\textwidth}{$^{2}$\textsl{\small Physics Department, Taras
Shevchenko Kiev University, 6 Academician
 Glushkov pr, Kiev 03127, Ukraine}}}

\date{}
\maketitle

\begin{abstract}Mathematical method of quantum phase space is very
useful in physical applications like quantum optics and
non-relativistic quantum mechanics. However, attempts to
generalize it for the relativistic case lead to some difficulties.
One of problems is band structure of energy spectrum for a
relativistic particle. This corresponds to an internal degree of
freedom, so-called charge variable. In physical problems we often
deal with such of dynamical variables that do not depend on this
degree of freedom. These are position, momentum, and any
combination of them. Restricting our consideration to this kind of
observables we propose the relativistic Weyl--Wigner--Moyal
formalism that contains some surprising differences from its
non-relativistic counterpart.
\end{abstract}


This paper is devoted to the phase space formalism that is
specific representation of quantum mechanics. This representation
is very close to classical mechanics and its basic idea is a
description of quantum observables by means of functions in phase
space (symbols) instead of operators in the Hilbert space of
states.

The first idea about this representation has been proposed in the
early days of quantum mechanics in the well-known Weyl work
\cite{Semenov:weyl}. Let us put the operator $\hat{A}$, which acts
in the Hilbert space of states of a quantum system, into
correspondence to a function on the phase space (symbol)
$A\left(p,q\right)$ according with the following rule
\begin{equation}
\hat{A}=\int\limits_{-\infty}^{+\infty}A\left(p,q\right)\hat{W}\left(p,q\right)
d p d q , \label{Semenov:equation1}
\end{equation}
where $\hat{W}\left(p,q\right)$ is an operator generalization of
$\delta$-function that is called operator of quasiprobability
density. This operator is defined as Fourier image of operator
exponent (operator of representation of the Heisenberg-Weyl group)
\begin{equation}
\hat{W}\left(p,q\right)=\frac{1}{2\pi\hbar}\int\limits_{-\infty}
^{+\infty}\exp\left\{\frac{i}{\hbar}\left[Q\left(p-\hat{p}\right)-
P\left(q-\hat{q}\right)\right]\right\}d Q d P.
\label{Semenov:equation2}
\end{equation}
Nowadays, this transformation is known as Weyl transform.

In table \ref{Semenov:table1} one can find correspondences among
some constructions in classical mechanics and in the Hilbert space
and phase space representations of quantum mechanics. First of
all, it is related to two binar operations, namely, usual product
and bracket. In classical mechanics there exist conventional
commutative multiplication of two functions and Poisson bracket.
It is well known that corresponding operations in the Hilbert
space representation of quantum mechanics are non-commutative
product of two observables and commutator. In the phase space
representation of quantum mechanics we deal with so-called star
product and Moyal bracket.

\begin{table}[ht]
\caption{\label{Semenov:table1} Some constructions in classical
mechanics and in quantum mechanics (Hilbert space and phase space
representations) are shown. First line corresponds to the usual
product of two observables. Brackets (Poisson, commutator, Moyal)
are shown in the second line. Evolution equations in Heisenberg
and Schr\"odinger representations are presented in third and
fourth lines respectively. In the fifth line one can see
expressions for mean value of an observable. Expressions for a
pure state are presented in the sixth line.}
\begin{tabular}{|@{}>{\PBS\centering\hspace{0pt}}m{\textwidth/3-4pt}@{}|
@{}>{\PBS\centering\hspace{0pt}}m{\textwidth/3-4pt}@{}|
@{}>{\PBS\centering\hspace{0pt}}m{\textwidth/3+5pt}@{}|} \hline
Classical mechanics & Quantum mechanics\newline (Hilbert space) &
Quantum mechanics\newline (phase space)\\ \hline\rule{0pt}{0pt}
$A\left(p,q\right)B\left(p,q\right)$ &
\rule{0pt}{0pt}$\hat{A}\hat{B}$ &
\rule{0pt}{0pt}$\scriptstyle{A\left(p,q\right)\star
B\left(p,q\right)},$
\newline $\scriptstyle{\star=\exp\left\{\frac{i\hbar}{2}\left(
\overleftarrow{\partial_q}\overrightarrow{\partial_p}-
\overleftarrow{\partial_p}\overrightarrow{\partial_q}\right)\right\}}$
\\ \hline
\rule{0pt}{0pt}$\begin{array}{l}\left\{A(p,q),B(p,q)\right\}_P
\\
\scriptstyle{=A(p,q)\left(
\overleftarrow{\partial_q}\overrightarrow{\partial_p}-
\overleftarrow{\partial_p}\overrightarrow{\partial_q}\right)B(p,q)}\end{array}$&
\rule{0pt}{0pt}$\left[\hat{A},\hat{B}\right]=\hat{A}\hat{B}-\hat{B}\hat{A}$&
\rule{0pt}{0pt}$\begin{array}{l}\left\{A(p,q),B(p,q)\right\}_M
\\ \scriptstyle{= \frac{1}{i\hbar}\left(A(p,q)\star B(p,q)-B(p,q)\star
A(p,q)\right)}\\
\scriptstyle{=\frac{2}{\hbar}A(p,q)\sin\left\{\frac{\hbar}{2}\left(
\overleftarrow{\partial_q}\overrightarrow{\partial_p}-
\overleftarrow{\partial_p}\overrightarrow{\partial_q}\right)\right\}B(p,q)}\end{array}$\\
\hline \rule{0pt}{0pt}$\partial_tA=\left\{A,H\right\}_P$&
\rule{0pt}{0pt}\newline$\partial_t\hat{A}=\frac{1}{i\hbar}\left[\hat{A},\hat{H}\right]$
\rule{0pt}{0pt}\newline&\rule{0pt}{0pt}$\partial_tA=\left\{A,H\right\}_M$\\
\hline\rule{0pt}{0pt}$\partial_t\rho=\left\{H,\rho\right\}_P$&
\rule{0pt}{0pt}\newline$\partial_t\hat{\rho}=\frac{1}{i\hbar}\left[\hat{H},\hat{\rho}\right]$
\newline&\rule{0pt}{0pt}$\partial_tW=\left\{H,W\right\}_M$\\ \hline
\rule{0pt}{0pt}$\scriptstyle{\bar{A}=\int\limits_{-\infty}^{+\infty}A\left(p,q\right)
\rho\left(p,q\right)d p d q}$
&\rule{0pt}{0pt}\newline$\bar{A}=\Tr\left(\hat{A}
\hat{\rho}\right)$\newline&\rule{0pt}{0pt}$\scriptstyle{\bar{A}=\int\limits_{-\infty}^{+\infty}
A\left(p,q\right)W\left(p,q\right)d p d q}$
\\ \hline
\rule{0pt}{0pt}$\begin{array}{l}\scriptstyle{\rho\left(p,q\right)
=\delta\left(p-p(t)\right)\delta\left(q-q(t)\right)}\\
\scriptstyle{}\end{array}$&
\rule{0pt}{0pt}$\hat{\rho}=\left|\psi\right\rangle\left\langle\psi\right|$&
\rule{0pt}{0pt}$\begin{array}{l}\scriptstyle{W\left(p,q\right)}
\\\scriptstyle{= \frac{1}{{2\pi \hbar}}\int\limits_{ - \infty }^{ + \infty }
{\psi^ *  \left( {p +\frac{P}{2}} \right)\psi \left( {p -
\frac{P}{2}} \right)e^{ - \frac{i}{\hbar }Pq}dP }}\end{array}$\\
\hline
\end{tabular}

\end{table}

One can see a very important property: when $\hbar\rightarrow0$,
quantum mechanics in the phase space transforms to classical
mechanics.

Now, it is possible to write down quantum counterparts for the
evolution equations (Hamilton or Liouville), expressions for
expected values of observables, and distribution function for a
pure state. In quantum mechanics this function has been presented
firstly by Wigner in 1932 \cite{Semenov:wigner}, and nowadays is
well-known as Wigner function.

This formalism is very useful in a lot of branches of modern
physics, {\em e.g.} in quantum optics (see for an example
\cite{Semenov:schleich}.) The fact that Wigner function does not
satisfy Kolmogorov axiomatic for probability has helped us to
better understand the nature of quantum non-locality
\cite{Semenov:klyshko}, \cite{Semenov:vogel}.

However, generalization of this formalism to the relativistic case
meets some problems. A very interesting feature is the fact that
Weyl transform is not Lorentz invariant in an explicit form. We do
not consider this question in the paper. Some consideration of it
and references one can find in \cite{Semenov:lev1},
\cite{Semenov:lev2}. Here, we concentrate our attention on another
problem, that is absence of a well-defined position operator in
relativistic case.

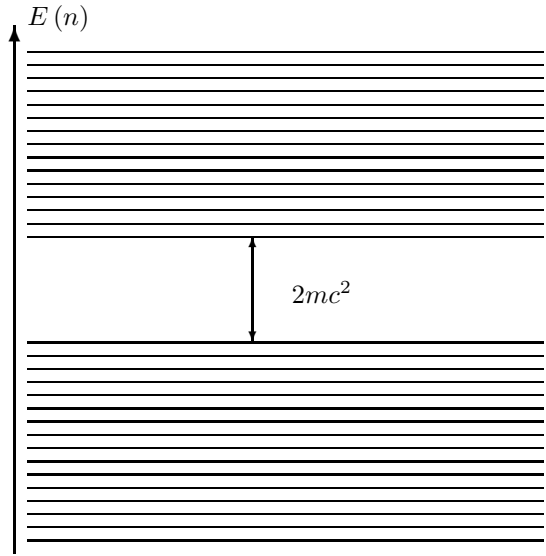
\begin{figure}[ht]
\begin{center}\parbox{200pt}{\begin{picture}(200,200)
\put(0,0){\thicklines\vector(0,1){200}}
\put(5,200){$E\left(n\right)$}
\multiput(5,80)(0,-5){16}{\line(1,0){195}}
\multiput(5,120)(0,5){15}{\line(1,0){195}}
\put(90,80){\vector(0,1){40}} \put(90,120){\vector(0,-1){40}}
\put(105,95){$2mc^2$}
\end{picture}}\end{center}
\caption{\label{Semenov:figure1} Energy spectrum of a relativistic
particle. }
\end{figure}

Consider the simplest example, a scalar charged particle in a
constant magnetic field. This system is described by the
Klein--Gordon equation
\begin{equation}
-\hbar^2\partial_t^2\Psi=
\left(c^2\left(\hat{p}-eA\left(q\right)\right)^2+m^2c^4\right)\Psi.\label{semenov:equation3}
\end{equation}
This is well-known that energy spectrum of this equations is
subdivided in two bands: lower and upper (see Fig.
\ref{Semenov:figure1})
\begin{equation}
E\left(n,\pm\right)=\pm
mc^2\sqrt{1+\frac{2}{mc^2}e\left(n\right)},
\label{Semenov:equation4}
\end{equation}
where $e\left(n\right)$ is the spectrum of a non-relativistic
particle. Moreover, according with modern notions about the field
theory, one supposes that lower band is occupied by particles
(so-called Dirak Sea.)

The Klein--Gordon equation is a second-order equation in time. One
can rewrite it in accordance with the Feshbach--Villars formalism
\cite{Semenov:feshbach} as a system of two first-order equations.
To achieve this one should provide changing variables
\begin{eqnarray}
\Psi=\frac{1}{\sqrt{2}}\left(\varphi+\chi\right),&
i\hbar\partial_t\Psi=\frac{mc^2}{\sqrt{2}}\left(\varphi-\chi\right).\label{Semenov:equation6}
\end{eqnarray}
As a result we can write down the Klein--Gordon equation in the
form of the Schr\"odinger equation
\begin{equation}
i\hbar\partial_t\psi=\hat{H}\psi, \label{Semenov:equation7}
\end{equation}
with two-component wave function
$\psi=\left(\begin{array}{c}\varphi\\ \chi\end{array}\right)$ and
Hamiltonian
$\hat{H}=\left(\tau_3+i\tau_2\right)e\left(\hat{n}\right)+ \tau_3
mc^2$. In the last equation $e\left(\hat{n}\right)$ is
non-relativistic Hamiltonian, and $\tau_i$ are the Pauli matrices.
In fact, they present a specific internal degree of freedom,
so-called charge variable.

The point is that eigenfunctions of position and momentum
operators belong to both bands. Hence, the question about
measurement of position and momentum in the relativistic case is
not so trivial.

If one has an aim to construct the phase space formalism for a
relativistic particle, the charge variable should be taken into
account as an independent degree of freedom, and $\tau_i$ have to
be considered as a specific part of the phase space. Such a kind
of formalism would be very useful. However, we will consider a
simpler case.

Let us restrict our attention only to such observables that
commute with all matrices $\tau_i$
\begin{eqnarray}
\left[\hat{A},\tau_i\right]=0,& \textrm{for}& i=1,2,3.
\label{Semenov:equation10}
\end{eqnarray}
In fact, this means that these observables are operator functions
of position and momentum and do not depend on $\tau_i$
\begin{equation}
\hat{A}= F\left(\hat{p},\hat{q}\right).\label{Semenov:equation11}
\end{equation}

We call elements of corresponding subalgebra of the dynamical
algebra as {\em charge-invariant observables}. Indeed, all these
operators are invariant relatively to unitary transformations (or,
to be more precise, generalized unitary transformations
\cite{Semenov:feshbach}) in charge subspace of the Hilbert space.
A lot of important observables belong to the subalgebra of
charge-invariant observables. These are position, momentum, second
moments, {\em etc}. However, it is worth noting that Hamiltonian
and current do not belong to this subalgebra. Hence, one cannot
use formalism presented here for those ones.

Using the phase space formalism, one can provide analysis for such
a subalgebra of dynamical algebra. An interesting feature of these
observables is the fact that even $[A]$ and odd $\{A\}$ parts of
corresponding operators (diagonal and not diagonal components of
matrices $2\times2$ in the energy representation, see
\cite{Semenov:feshbach}) are not independent. In
\cite{Semenov:lev1}, \cite{Semenov:lev2}, one can find a proof of
the following constraint for them (in the energy representation)
\begin{equation}
\{A\}_{nm}=\frac{{E(m) - E(n)}}{{E(m) + E(n)}} \tau_{1}
[A]_{nm}.\label{Semenov:equation12}
\end{equation}

Another very important property of charge-invariant observables is
the fact that it is possible for them to present usual (scalar,
not matrix-valued) Wigner function with conventional rule for the
calculation of mean values
\begin{equation}
\overline{A}=\int\limits_{-\infty}^{+\infty}A(p,q)W(p,q)dpdq.
\label{Semenov:equation13}
\end{equation}

Nevertheless, there are very important differences from the
non-relativistic case. First of all, this object is a sum of four
components
\begin{equation}
W(p,q)=\sum_{\alpha=\pm 1
}\left(W_{[\alpha]}(p,q)+W_{\{\alpha\}}(p,q)\right).\label{Semenov:equation14}
\end{equation}
Namely, these are two even components, which correspond to
particle and antiparticle
\begin{equation}
W_{[\pm]}(p,q) =\!\sum_{n,m=0}^{\infty}{\varepsilon(m,n)} W_{nm}
(p,q)C_{m;}^*{}_{\pm}C_{n;}{}^\pm,\label{Semenov:equation15}
\end{equation}
and two odd (non-physical) components, which correspond to the
interference between particle and antiparticle
\begin{equation}
W_{\{\pm\}}(p,q) =\!\sum_{n,m=0}^{\infty}{\chi(m,n)} W_{nm}
(p,q)C_{m;}^*{}_{\pm}C_{n;}{}^{\mp}.\label{Semenov:equation16}
\end{equation}

In equations (\ref{Semenov:equation15}) and
(\ref{Semenov:equation16}) $C_{n;}{}^\pm$ is a wave function in
the energy representation, $W_{nm} (p,q)$ is a Hermitian
generalization of the Wigner function (see \cite{Semenov:lev2},
\cite{Semenov:fairlie}), functions
\begin{equation}
\varepsilon \left(n,m\right) = \frac{{E\left(n\right) +
E\left(m\right)}}{{2\sqrt
{E\left(n\right)E\left(m\right)}}}\label{Semenov:equation17}
\end{equation}
and
\begin{equation}
\chi \left(n,m\right) = \frac{{E\left(n\right) -
E\left(m\right)}}{{2\sqrt
{E\left(n\right)E\left(m\right)}}}\label{Semenov:equation18}
\end{equation}
are so-called $\varepsilon$- and $\chi$-factors.

One can find evolution equations (quantum Liouville equation) for
each component separately
\begin{eqnarray}
\partial_tW_{[\pm]}(p,q,t)=\pm\left\{E(p,q),W_{[\pm]}(p,q,t)\right\}_M,\label{Semenov:equation19} \\
\partial_tW_{\{\pm\}}(p,q,t)=\mp\left[E(p,q),W_{\{\pm\}}(p,q,t)\right]_M,\label{Semenov:equation20}
\end{eqnarray}
where
\begin{equation}
E(p,q) = \sqrt[\star]{m^2 c^4  + c^2
\left(p-eA\left(q\right)\right)^2}\label{Semenov:equation21}
\end{equation}
is the effective Hamiltonian. Here one has two surprising features
as well.

First of all, the symbol of the Hamiltonian is defined by means of
square root in a sense of star product. This means that Weyl
symbol of the Hamiltonian differs from the Hamilton function in
classical mechanics. In non-relativistic mechanics they coincide.

Second, the evolution equation for the odd components is
formulated using ``anti-Moyal'' bracket, which is counterpart of
the anti-commutator. This means that evolution of odd part is a
non-unitary operation.

Moreover, there are once more, the most important property.
Definition of even component include a very important multiplier,
$\varepsilon$-factor (\ref{Semenov:equation17}). It is absent in
the non-relativistic theory. Consider its properties more closely:
\begin{enumerate}
\item $\varepsilon(n,m)=\varepsilon(m,n)$.
\item $\varepsilon(n,n)=1$.
\item $\varepsilon(n,m)>1$, if $n
\neq m$.
\end{enumerate}

This is very similar to the case, when one considers an arbitrary
system in an environment. For the Wigner function in this case one
can write down the following expression
\begin{equation}
W(p,q) =\sum\limits_{n=0}^{\infty} W_{nn}
(p,q)\left|C_{n}\right|^2+\sum\limits_{n\neq m}a(m,n) W_{nm}
(p,q)C_{m}^*C_{n}.\label{Semenov:equation22}
\end{equation}
However, non-diagonal components of $a(m,n)$ (counterpart of
$\varepsilon$-factor) are always smaller than~$1$. Therefore, in
the relativistic case, influence of vacuum is inverted to the
decoherence process. This means that relativistic system has more
quantum non-local properties than its non-relativistic
counterpart.

It is possible to consider this question from the more formal
point of view. It is well-known that not every normalized function
can be regarded as Wigner function presenting a real quantum
physical state \cite{Semenov:tatarskii}. In the non-relativistic
case there exists so-called quantization condition that selects
Wigner functions from set of functions on the phase space. In our
case this condition can be presented in the following way (for a
free particle).
\begin{theorem}\label{Semenov:theorem1}
For the functions $W_{[\pm]}\left(p,q\right)$ and
$W_{\{\pm\}}\left(p,q\right)$ to be even and odd components of the
Wigner function for charge-invariant observables, it is necessary
and sufficient that all constrains hold true (see
\cite{Semenov:lev1}) and the following conditions are satisfied:
\begin{eqnarray}
\frac{{\partial ^2 }}{{\partial p_1 \partial p_2 }}\ln
\int\limits_{ - \infty }^{ + \infty }
W_{[\pm]}\left(\frac{1}{2}(p_1 + p_2
),q\right)\exp\left(\frac{i}{\hbar }(p_1  - p_2 )q\right) dq
\nonumber\\ =- \frac{c^4 p_1 p_2 }{E(p_1 )E(p_2 )(E(p_1 ) + E(p_2
))^2 },\label{Semenov:equation23}
\end{eqnarray}
\begin{eqnarray}
\frac{{\partial ^2 }}{{\partial p_1 \partial p_2 }}\ln
\int\limits_{ - \infty }^{ + \infty }
W_{\{\pm\}}\left(\frac{1}{2}(p_1 + p_2
),q\right)\exp\left(\frac{i}{\hbar }(p_1  - p_2 )q\right) dq
\nonumber\\ =- \frac{c^4 p_1 p_2 }{E(p_1 )E(p_2 )(E(p_1 ) - E(p_2
))^2 }.\label{Semenov:equation24}
\end{eqnarray}
\end{theorem}
A proof of this criterion can be found in \cite{Semenov:lev1}, and
formulation for a more general case in \cite{Semenov:lev2}.

The main difference from the non-relativistic case is the fact
that in the right-hand-side of equation (\ref{Semenov:equation23})
one has an expression that differs from $0$. In the
non-relativistic case this is exactly $0$.

\begin{figure}[ht]
\begin{center}
\includegraphics[width=0.49\textwidth, clip=]{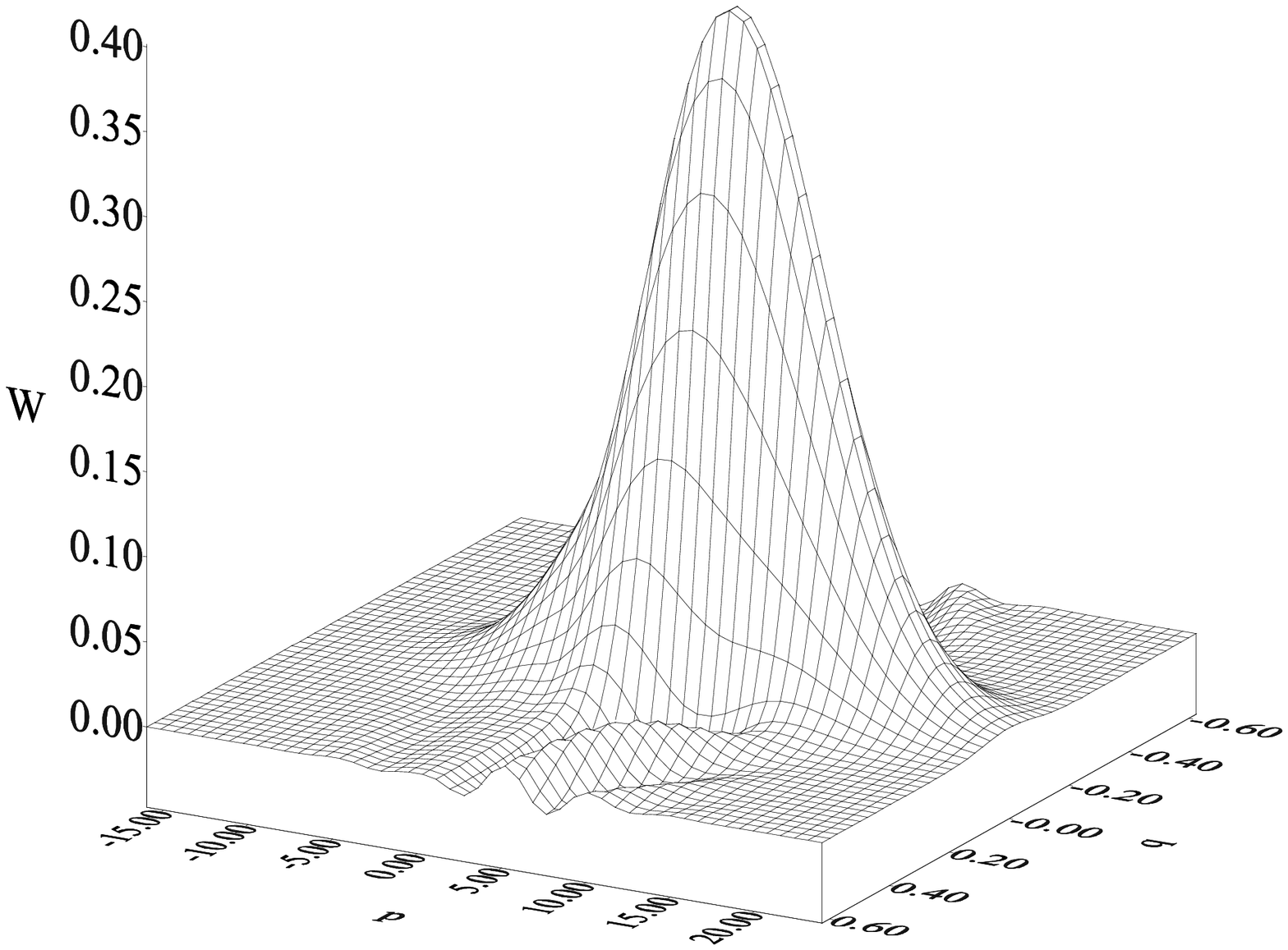}
\includegraphics[width=0.49\textwidth, clip=]{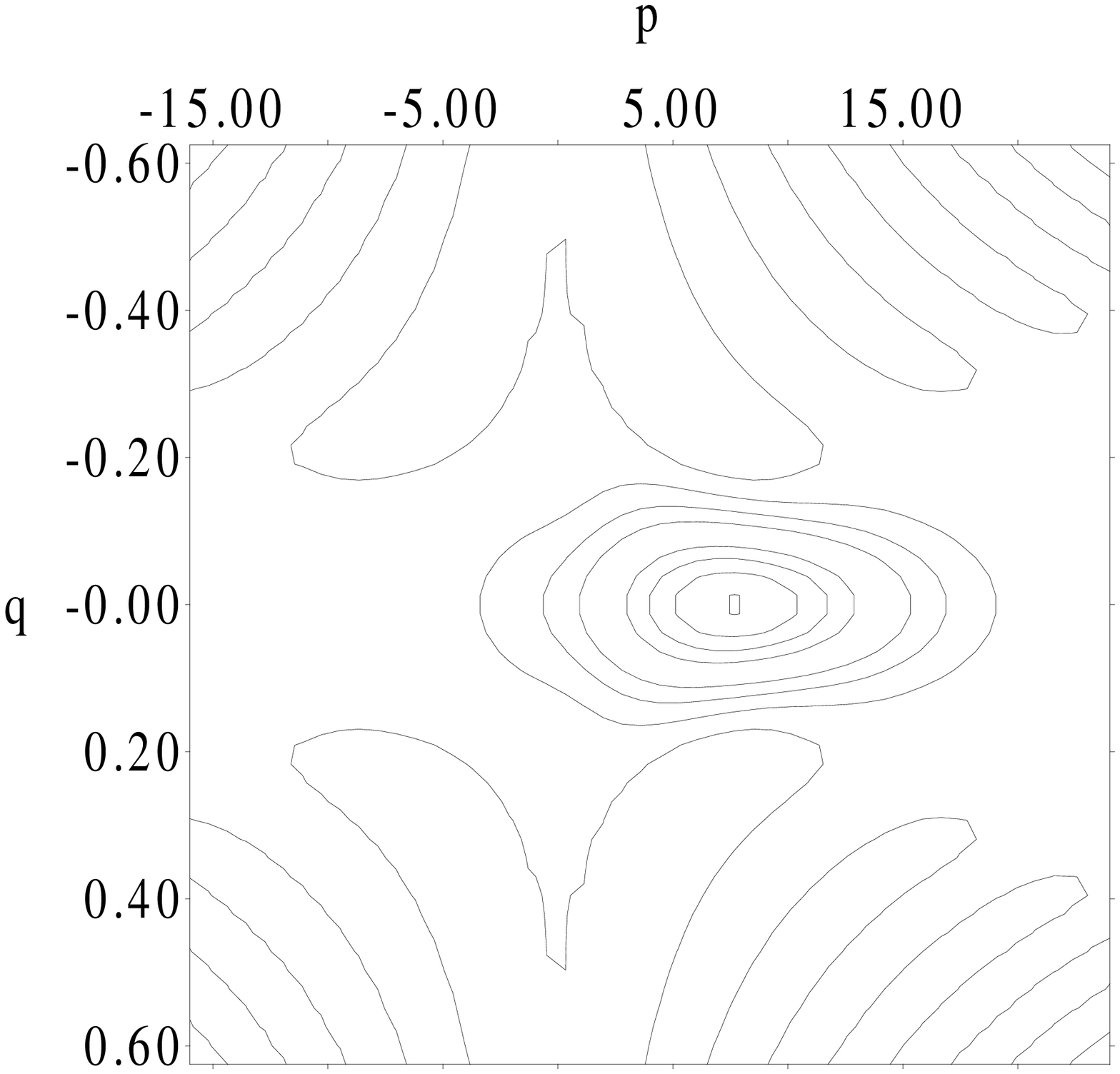}
\end{center}
\caption{\label{Semenov:figure2} Wigner function and its contours
for the coherent state of a free relativistic particle. Position
is given in the units of Compton wavelength, momentum is given in
$mc$ units.}
\end{figure}

Consider a simple example. Figure \ref{Semenov:figure2} shows the
Wigner function for the coherent state of a free particle
\cite{Semenov:lev3}. In fact, this is a Gauss distribution for
non-strong space localization . However, when localization along
position is very strong, additional ``vacuum fluctuations''
appear. They counteract to the strong localization. This means
that coherent state (which is classical in terms of work
\cite{Semenov:vogel}) manifests a quantum feature.

For a conclusion, we note that quantum quasidistributions of
relativistic systems are more different from classical
distributions than their non-relativistic counterparts. This means
that relativistic systems are more interesting objects for
investigation of quantum non-locality and non-classicality.

\subsection*{Acknowledgements}

The authors are very grateful to the organizers of the Fifth
International Conference ``Symmetry in Nonlinear Mathematical
Physics'' for the opportunity to present this work.

\end{document}